# Spontaneous Formation of Lateral Defect-Density Gradient in CVD-Grown $MoS_2$ Monolayers

Marek Kostka[1,2,3], Hsin-Sung Chen[3], Cheng-Ying Hsiao [3], Yan-Ru Chu[3], Jindřich Mach[1,2], Miroslav Bartošík[1,2], Jakub Šťastný[1], Tomáš Šikola[12], Wen Chin Lin[3]

[1] Institute of Physical Engineering, Brno University of Technology, Technická 2896/2, 616 69 Brno, Czech Republic,

[2] Central European Institute of Technology (CEITEC), Purkyňova 656/123, 612 00 Brno, Czech Republic

[3] Department of Physics, National Taiwan Normal University, Taipei, Taiwan

Abstract:

*Defect engineering in two-dimensional materials is a powerful strategy for tuning electronic, optical, and catalytic properties. While pristine molybdenum disulfide ($MoS_2$) has been thoroughly studied, defective $MoS_2$ remains poorly understood due to the complexity and inconsistency of controlled defect creation. Here, we report a defect-density gradient that forms spontaneously in $MoS_2$ flakes during chemical vapor deposition (CVD) growth. Photoluminescence and Raman spectroscopy mapping reveal the lateral arrangement of this gradient across a single flake. Deconvolution of the excitonic photoluminescence emission identifies electron doping as the key factor governing the spatial variation in optical response. This doping likely originates from sulfur vacancies formed during CVD growth, as further supported by the observed Raman mode shifts. We propose that this gradient arises from precursor depletion near the nucleation center during growth, resulting in a gradual variation in defect density. This lateral defect structure provides a convenient platform for systematically characterizing defective $MoS_2$ across a continuous range of defect densities within a single sample. The sulfur-vacancy-driven surface activation identified here can also be important for future catalytic applications.*

## Introduction:

Molybdenum disulfide ($MoS_2$) is recognised as a promising two-dimensional semiconductor with remarkable optical and electronic properties. In its monolayer form, $MoS_2$ exhibits a direct bandgap and strong photoluminescence (PL), in contrast to the indirect bandgap of bulk and multilayer $MoS_2$ [1]. Reduced dielectric screening and strong Coulomb interactions in the two-dimensional limit [2] result in large exciton binding energies, allowing excitonic states to remain stable at room temperature. Its two-dimensional nature enables its electronic and optical properties to be tuned through strain [3–7], doping [8–10] or defect engineering [11,12]. Although pristine $MoS_2$ is generally preferred for studies of fundamental light-matter interactions, defect engineering offers an alternative strategy for tailoring the material towards applications such as broadband photodetectors [13,14], gas sensors [15,16], and hydrogen production [17,18].

Common preparation methods for $MoS_2$ include mechanical exfoliation (ME) and chemical vapour deposition (CVD). While ME produces high-quality crystals, it offers limited efficiency, scalability and reproducibility. In contrast, CVD enables large-area growth of $MoS_2$ but typically introduces defects that reduce the crystal quality. Sulfur vacancies ($V_S$) are the dominant defect type in CVD-grown $MoS_2$, whereas other defects, including molybdenum vacancies ($V_{Mo}$) and antisites ($S_{Mo}$ and $Mo_S$), are less prevalent [11]. Importantly, $V_S$ can substantially alter the local electronic structure and chemical reactivity of $MoS_2$. Vacancy-induced electronic states can modify charge-carrier [19] and optical [20] behaviour, while the altered local bonding around sulfur vacancies can promote molecular adsorption and increase chemical reactivity [21–23]. Consequently, engineering the concentration and distribution of $V_S$ provides a route towards tailoring $MoS_2$ for sensing and catalytic applications, including the hydrogen evolution reaction. Numerous approaches have therefore been developed to introduce defects in $MoS_2$. Most rely on external stimuli to control defect formation, including plasma exposure [24], ion bombardment [25,26] and gas annealing [23,27]. While these methods allow the defect density to be modified after growth, they require additional processing and may introduce unwanted structural modifications. An alternative strategy is to engineer defects directly during material synthesis. Spontaneous defect formation during growth eliminates the need for additional chemical or physical treatment [28]. Moreover, CVD offers a scalable route towards the production of defective $MoS_2$. Direct control of defect formation during CVD could therefore combine material synthesis and defect engineering within a single scalable process.

Most studies, however, focus on layers with spatially uniform defect distributions. Such samples provide information only at a limited range of defect densities, making systematic defect-dependent studies reliant on comparisons between separate samples. Variations between samples can then complicate the direct correlation between defect density and the resulting material properties. This limitation can be addressed by introducing a non-uniform defect distribution across the $MoS_2$ surface. A continuous defect-density gradient provides a range of defect concentration within a single monolayer, enabling defect-dependent changes in material properties to be investigated while reducing sample-to-sample variability. This concept is particularly compatible with spatially resolved spectroscopic techniques, such as PL and Raman mapping, which can probe the optical and vibrational response continuously across the same flake. Beyond its use as an experimental platform, spatial control over $V_S$ may provide a route towards laterally varying optical, electronic, or chemical functionality within a chemically homogeneous $MoS_2$ monolayer. Such defect-defined structures could potentially enable spatially patterned sensing or catalytic activity without introducing a second material or chemical composition. Previous studies of lateral heterostructures have primarily relied on variations in chemical composition [17,29–31], whereas the direct formation of a lateral defect-density gradient in a single growth step has not yet been demonstrated experimentally.

In this work, we present a spontaneously formed gradient of defect density within single-layer $MoS_2$ flakes. The continuous transition from pristine to defective monolayer $MoS_2$ originates directly from the CVD growth process. Using spatially resolved PL and Raman spectroscopy, we map the optical and vibrational response associated with increasing defect density across individual flakes. Our results demonstrate that spatially varying defect densities can be generated directly during $MoS_2$ growth and investigated within a single continuous monolayer, providing a scalable platform for studying defect-dependent properties and a potential route towards spatially engineered functionality in two-dimensional materials.

Results:

The $MoS_2$ flake was first characterized using microscopic techniques. Fig. 1a) shows an optical microscope image of the $MoS_2$ flake, with no visible inhomogeneity on the surface. The SEM image in Fig. 1b), obtained using secondary electron signal, also shows no contrast on the flake. Fig. 1c) shows the morphology of the $MoS_2$ flake in AFM, with focus on the flake edge emphasized in Fig. 1b by white square. The height profile in the inset plot (along the black line in Fig. 1c) reveals the difference of height $\Delta h = 0.8\ nm$, which confirms a single-layer nature of $MoS_2$. Therefore, the flake is homogeneous, and any observed differences must be intrinsic, rather than being caused by thickness variance.

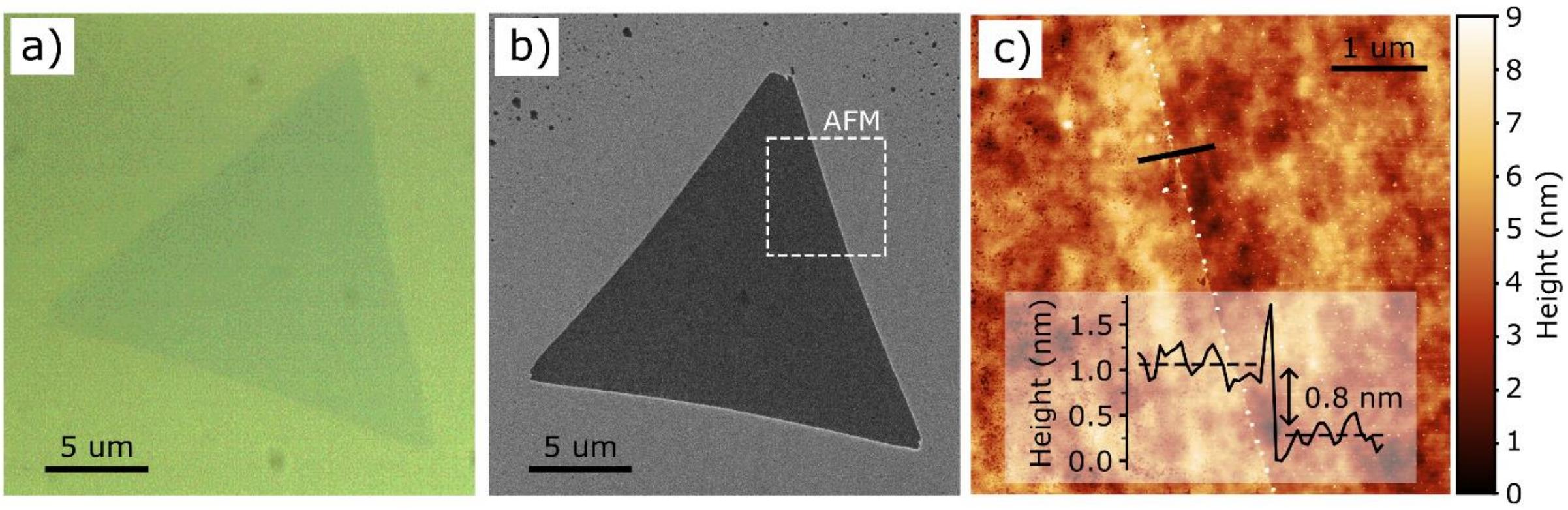

*Fig. 1: Triangular $MoS_2$ flake grown by CVD. a) Optical microscope image, b) SEM image showing the contrast in secondary electrons, c) AFM topography image with height profile across the flake edge, indicating the flake is a single layer $MoS_2$.*

Photoluminescence (PL) mapping of the $MoS_2$ flake first revealed an inhomogeneity in the $MoS_2$ quality. To analyse the PL data correctly, the PL peak of $MoS_2$ was deconvolved into three most prominent components: A exciton, $A^-$ trion and B exciton (as illustrated in Fig. 2d), e)). The exciton energies are defined by the energy levels within the valence band, which are split due to strong spin-orbit coupling. This gives the exciton quasiparticles a large binding energy of $E_b \approx 140$ meV, making them stable at room temperature. The $A^-$ trion is a negatively charged quasi-particle that arises from an additional electron binding to the neutral A exciton. The $A^-$ trion binding energy is estimated from the energy separation between the $A^-$ trion and the A exciton as $E_{b,A^-} \approx 40$ meV. The main component of the PL emission is the A exciton. Fig. 2a) shows the intensity of the A exciton mapped over the flake. There is a gradual decrease in intensity in the bottom region, with a continuous gradient over a relatively short distance of 15 μm. Such spatial variation is uncommon, as most $MoS_2$ flakes are homogeneous or exhibit edge-to-centre variation. The channel of PL emission can be studied by comparing the intensity of $A^-$ trion and A exciton. Fig. 2b) shows the trion/exciton intensity ratio, emphasizing that the PL is dominated by A exciton in the pristine region, while $A^-$ trions are dominant in the defective region. The energy of the A exciton is mapped in Fig. 2c). This demonstrates a red shift in the bottom region, indicating bandgap narrowing. Fig. 2d) and 2e) show characteristic PL spectra from opposite sides of the gradient (for pristine and defective region in Fig. 2d) and 2e), respectively), accompanied by an illustrative spectrum deconvolution into A exciton, B exciton and $A^-$ trion.

This observation can be explained using two phenomena: growth-induced strain or chemical inhomogeneity. The strain effect is commonly observed in CVD-grown $MoS_2$ flakes, primarily due to the contraction mismatch of $MoS_2$ and substrate during the cooling stage of the growth process. Adhesion to the substrate might induce tensile strain in the single layer, resulting in significant changes to the electronic properties and optical response of $MoS_2$. Such tensile strain would lead to bandgap narrowing or PL red shifting, similar to the presented data. In the literature, growth-induced strain is usually formed in the centre-to-edge geometry due to the contraction mismatch [32]. In our case, the strain would need to be induced in the bottom part of the flake, possibly due to stronger adhesion to the substrate. Thus, the growth-induced strain is not a key consideration in determining the origin of the gradient in this work.

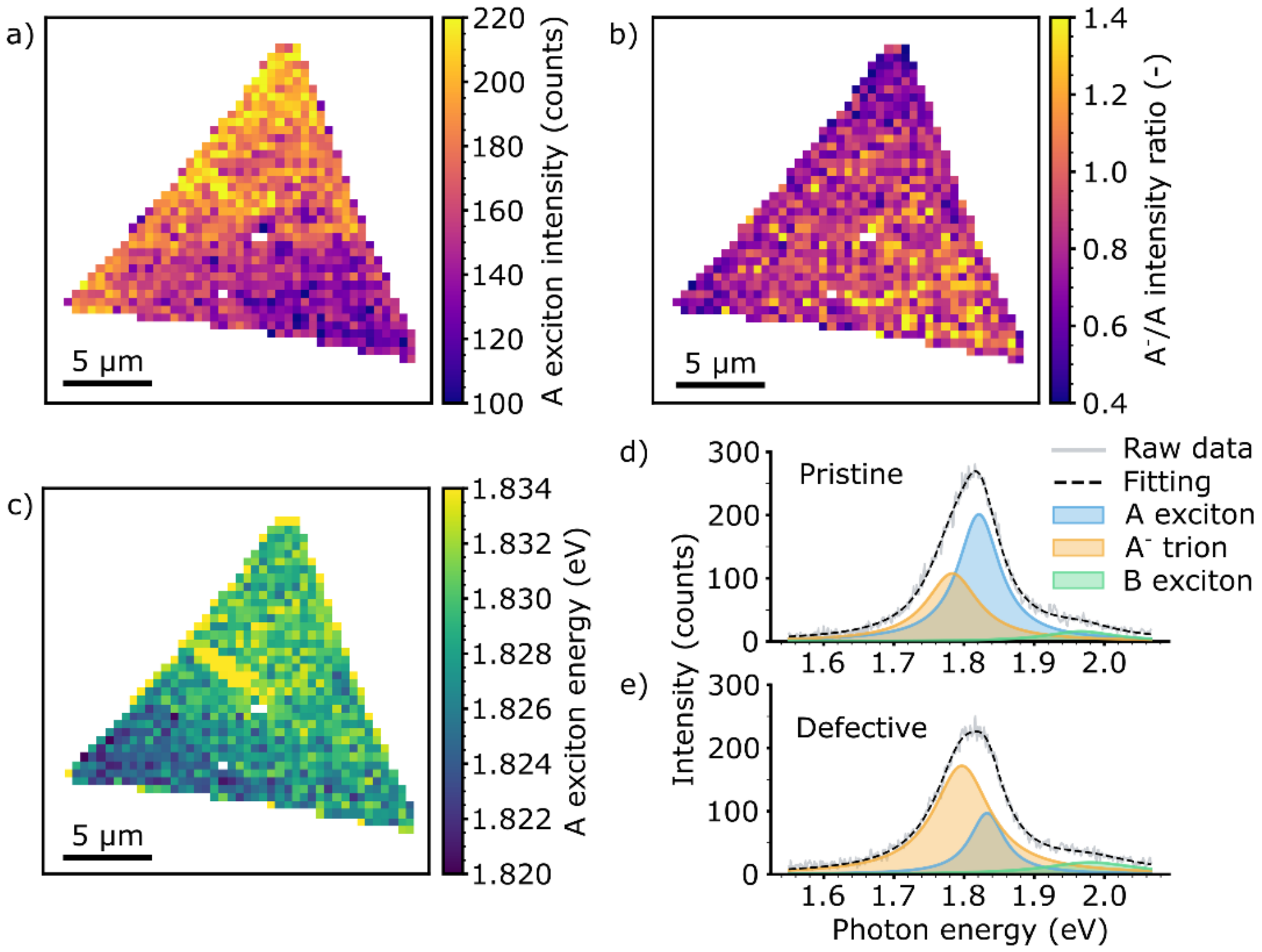

*Fig. 2: PL mapping of CVD-grown $MoS_2$ flake on a $SiO_2$ substrate. After deconvolution of the PL peak into A exciton, $A^-$ trion and B exciton, mapping of the A exciton shows a gradient of a) intensity and c) energy, illustrating the energy shift induced by electron doping. b) Trion/exciton intensity ratio indicating the main PL channel across the flake. d) PL spectrum taken from the pristine region with peak deconvolution. e) PL spectrum from the defective region to show the difference in PL peak composition. The rise of $A^-$ trion, in contrary to the decrease of A exciton, indicates a shift in the PL channel over the flake.*

A common distortion of the chemical homogeneity of $MoS_2$ is the formation of sulfur vacancies, which occurs particularly during growth. The growth mechanism is sensitive to the directionality and composition of the precursor flow, so any inhomogeneity in precursor flow (due to chamber geometry, sample attachment, etc.) results in a defective $MoS_2$ layer. As sulfur atoms are supplied insufficiently, the S:Mo ratio decreases and sulfur vacancies are created. They translate into electronic properties of $MoS_2$, because sulfur vacancies act as donors, thus inducing electron doping (n-doping) into the layer. The additional electrons bind to the neutral A excitons, which increases the population of charged $A^-$ trions. The A exciton population is reduced, decreasing its scattering rate and, subsequently, leading to PL quenching in the bottom region. The A exciton red shift is then due to one of three factors: i) local strain relaxation around sulfur vacancies, ii) doping-induced bandgap renormalisation, or iii) defect-bound exciton emission. Finally, $A^-$ trions decay non-radiatively due to the presence of sulfur vacancies. Proposed mechanisms for the channel of $A^-$ trion non-radiative decay are i) recombination centres on sulfur vacancies (via phonon generation) or ii) Auger recombination.

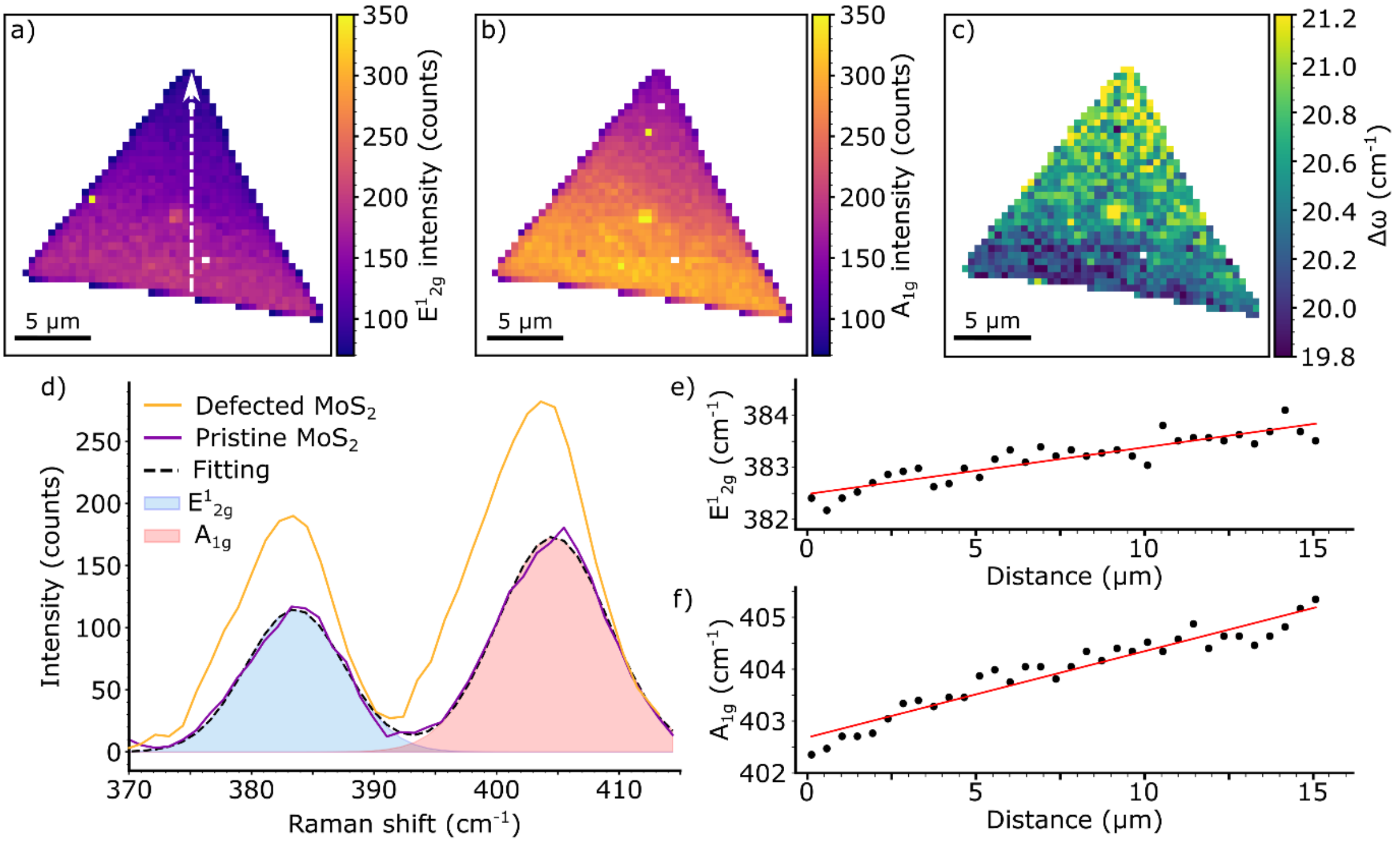

*Fig. 3: $MoS_2$ Raman spectroscopy mapping of the mode intensity, showing decrease in both a) in-plane $E^1_{2g}$ Raman mode and b) out-of-plane $A_{1g}$ mode. c) Raman spectroscopy mapping of the $E^1_{2g}$-$A_{1g}$ distance, reflecting the different trend between the two Raman modes. d) Single Raman spectra of two distinct regions of the flake. The Raman shift is extracted along the white line from a) to show the gradient of the Raman shifts of e) $E^1_{2g}$ mode and f) $A_{1g}$ mode. Their comparison shows that the $A_{1g}$ mode is more sensitive to the changes.*

The variation of PL signal can be correlated with Raman spectroscopy mapping of the $MoS_2$ flake, since lattice properties are sensitive to electronic changes and strain. Fig. 3a) and 3b) show maps of the $E^1_{2g}$ and $A_{1g}$ Raman mode intensities, picturing the gradient in the same direction, but with opposite slope compared to the PL mapping. The $A_{1g}$ optical phonon, which corresponds to the out-of-plane lattice vibration, exhibits a greater (more than twofold) increase in Raman scattering intensity in contrary to in-plane $E^1_{2g}$ mode. Fig. 3c) plots the distance between the Raman peaks of the $E^1_{2g}$ and $A_{1g}$ modes ($\Delta\omega = \omega(A_{1g}) - \omega(E^1_{2g})$). The gradient in this map indicates that the Raman mode shifts are affected differently, relative to the $MoS_2$ quality. Fig. 3d) shows two Raman spectra selected from the pristine and defective region, along with the fitted modes for illustration. The defective region shows stronger intensity than the pristine region. To describe the Raman shift gradient, Fig. 3e) and 3f) plot the extracted Raman shifts along the white line in Fig. 3 a) across the $MoS_2$ flake gradient. Both modes are fitted with a linear function, giving slopes of (0.041 ± 0.003) and (0.075 ± 0.004) $cm^{-1}/\mu m$ for the $E^1_{2g}$ and $A_{1g}$ modes, respectively.

The $A_{1g}$ Raman mode, with larger gradient across the flake, is sensitive to doping density in $MoS_2$, in contrary to the $E^1_{2g}$ mode which is not very affected by doping [5,33]. Therefore, the analysis of the Raman modes mapped across the flake indicates that electron doping is the primary cause of the gradient, particularly in the bottom region. Tensile strain would not affect the out-of-plane $A_{1g}$ mode to the extent [5] observed in Fig. 3b). Further evidence of electron doping can be found in the increased Raman intensity of both modes in the bottom region. As the electron density (and $A^-$ trion population) increases in the defective region, more phonons are generated in the scattering process, and a stronger Raman signal is observed in Fig. 3a) and 3b). Such Raman scattering enhancement contrasts with the PL quenching, but the increased electron density, originating from sulfur vacancies, suffices as an explanation for all the observed changes.

The observed Raman shifts of the $A_{1g}$ vibrational mode can be utilized for quantitative estimate of the doping density variation in $MoS_2$. Assuming a linear relationship $\Delta\omega = \alpha \cdot \Delta n$ with a proportionality coefficient of $\alpha = 2.22\ \mathrm{cm}^{-1}/10^{13}\mathrm{cm}^{-2}$ (based on literature reports [33,34]), the net change in doping density across the studied $MoS_2$ flake is estimated to $1.13 \times 10^{13} cm^{-2}$. This value agrees with the reported range of values for CVD-grown $MoS_2$ [11].

To confirm the origin of the observed gradient in the PL and Raman spectra of the $MoS_2$ flake, power-dependent PL and Raman spectroscopy (single spectra) measurements were performed in both the defective and pristine region. Fig. 4 shows the extracted peak intensities and energies (or Raman shifts), plotted against the excitation laser power in the range from 300 µW to 20 mW. The intensity and energy of the A exciton are plotted in Fig. 4a) and 4b), respectively, with intensity and laser power plotted on a logarithmic scale. The PL intensity should follow the power law $I \sim P^x$, where $x$ indicates the channel of energy dissipation [35]. The linear dependence in Fig. 4a) corresponds to $x$ close to unity, indicating radiative exciton recombination. The defective region (blue curve) shows a loss of linearity at higher powers (see inset in Fig. 4a), indicating exciton recombination saturation. This suggests a change in exciton behaviour with respect to defect density. In Fig. 4b), the A exciton energy shows a gradual red shift, as described elsewhere [36,37]. This is caused by an increase in the exciton population, which results in bandgap renormalisation. There is also a thermal effect due to lattice heating caused by the increased laser power. However, this trend changes in the defective region for optical powers above 1 mW (emphasized in green), where the red shift rate increases compared to pristine $MoS_2$. This might indicate a change in the energy dissipation channel. A suggested explanation is that the defect sites act as traps for photogenerated excitons. At low power, most of the excitons recombine radiatively without getting trapped. Above 1 mW of optical power, the defect traps become dominant and defect-mediated emission takes place, resulting in lattice heating. Such a change in local environment causes a bandgap renormalization, observed as an increase in red shift of the defective region.

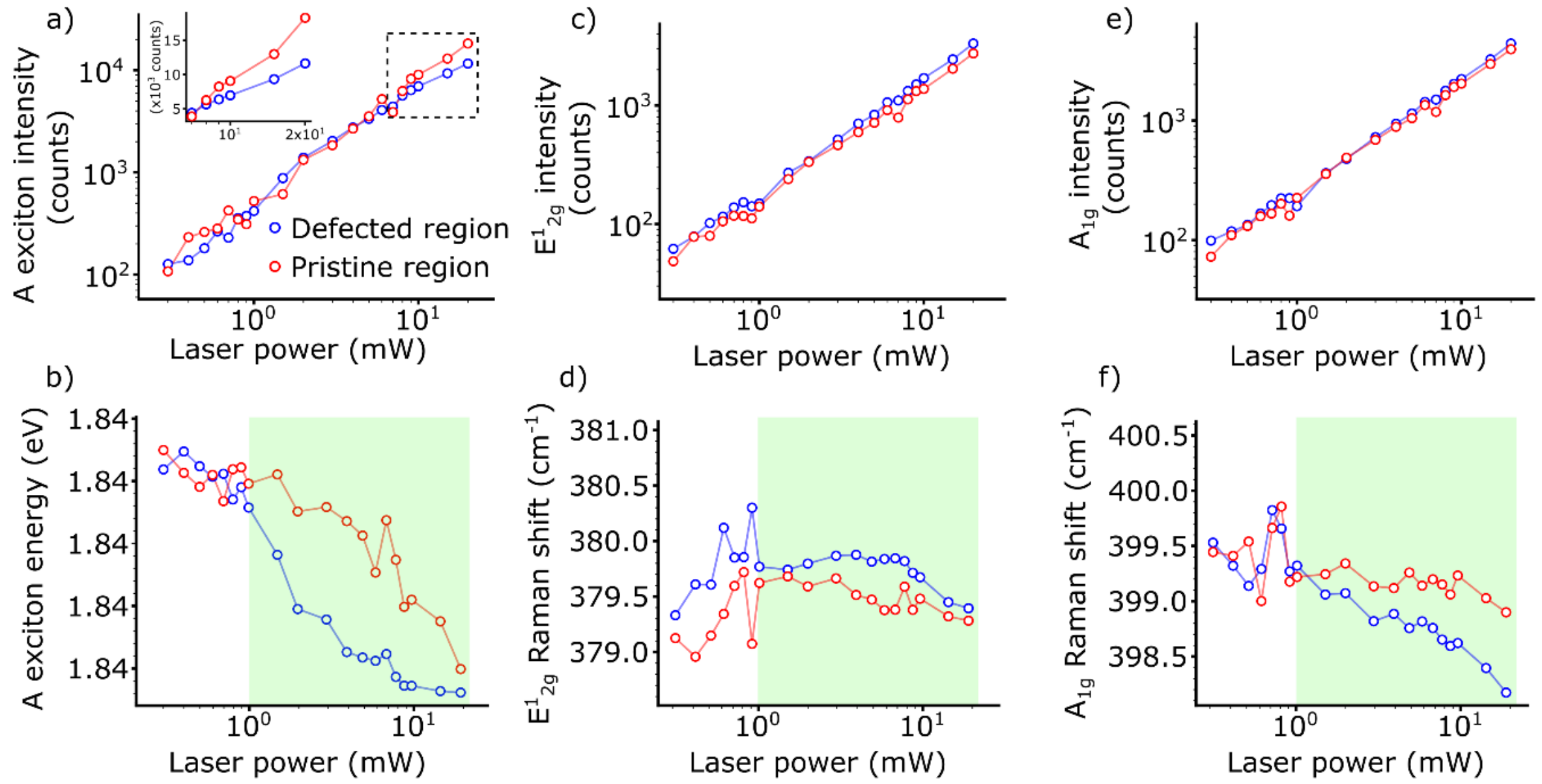

*Fig. 4: Power dependence of the optical properties of $MoS_2$ flakes, comparing the pristine (red) and defective (blue) region. Intensity and optical power are plotted on a logarithmic scale. The green background indicates the high-power regime. a) The intensity of A exciton shows a linear trend with a slope of unity, indicating the excitonic emission. b) The energy red shift of A exciton is caused by the bandgap renormalization arising from the increased exciton density. This trend*

*is, however, more pronounced in the defective region due to defect traps dominating energy dissipation, leading to lattice heating. c) and e) The intensity of $E^1_{2g}$ and $A_{1g}$ Raman modes in $MoS_2$ increases proportionally to the incident laser power. d) and f) The Raman shift of the $E^1_{2g}$ and $A_{1g}$ modes reflects the symmetry of the modes. The in-plane $E^1_{2g}$ mode is not greatly affected by defects, whereas the out-of-plane $A_{1g}$ mode changes significantly with heat dissipation saturation in defect sites at high powers (green background). This is evident as a change in slope in the doped region, in contrast to the more stable pristine region.*

Fig. 4c) – f) show the $E^1_{2g}$ and $A_{1g}$ Raman modes, along with their intensity and Raman shift dependencies on laser power. The intensities (Fig. 4c), 4e)) increase proportionally to the incident laser power, reflecting the increased phonon density. Defective region shows a subtle intensity increase compared to the pristine region, corresponding with the Raman intensity gradient in Fig. 3. Remarkably, Fig. 4d) and 4f) reflect a different nature of the $E^1_{2g}$ and $A_{1g}$ modes. The $E^1_{2g}$ peak is an in-plane vibrational mode and is not very sensitive to doping. Fig. 4d) shows a fluctuation in Raman shift with laser power; however, the difference in defect density only introduces a small constant offset in the Raman shift. In Fig. 4f), however, the power dependence of the $A_{1g}$ Raman shift exhibits two regimes (distinguished by green background). The first is a low-power regime in which the defective region remains unchanged from the pristine one. The second is a high-power regime (above 1 mW) in which the defective region exhibits a significant red shift in the Raman shift, notably deviating from the pristine region. This is consistent with the conclusion drawn from Fig. 4b) that energy is dissipated into the lattice. Such local lattice heating softens the bonds and thus decreases the Raman shift. Pristine $MoS_2$ shows a negligible shift in the $A_{1g}$ Raman shift with power, confirming that exciton recombination is sufficient. These observations confirm that sulfur vacancies (and corresponding electron density increase) are responsible for the change in optical properties of the $MoS_2$ flakes.

Regarding the origin of the sulfur vacancy-density gradient in $MoS_2$ flakes, this phenomenon can be attributed to (i) precursor depletion, (ii) local variations in the Mo:S ratio, or (iii) a temperature gradient. Although temperature can vary across the $SiO_2$ substrate, ohmic heating from the outer chamber walls is not expected to produce such spatially localized gradients within a single flake. Furthermore, if a temperature gradient were responsible, neighbouring flakes would be expected to show an even larger shift in PL energy relative to the flake under study – this was not observed. Local variation in the Mo:S ratio offers another possible explanation; however, a molybdenum excess would be expected to change flake morphology. While such morphological changes are indeed observed at the wafer scale (a transition from triangular to hexagonal flakes with dendritic structures), no comparable variation is seen at the microscopic scale within a single $MoS_2$ flake. This leaves precursor depletion as the most plausible explanation. As the gaseous sulfur precursor flows through the CVD chamber, it reaches the nucleation site via surface diffusion or direct adsorption, where it is progressively consumed during growth. This consumption leads to local sulfur depletion and, consequently, to the formation of sulfur vacancies in the direction of precursor flow. Supporting this interpretation, optical microscopy shows that upstream flakes are consistently larger than downstream ones (Supplementary Information, Fig. S1), consistent with a higher local precursor concentration upstream relative to downstream. Altogether, these observations demonstrate that localized precursor depletion during CVD growth is the primary cause of the sulfur vacancy-density gradient in $MoS_2$ flakes.

Conclusion:

In summary, we have demonstrated the spontaneous formation of a defect-density gradient in CVD-grown $MoS_2$ monolayer flakes. Spatially resolved photoluminescence and Raman spectroscopy mapping reveal how the optical properties vary with defect density across a single flake. Detailed analysis of the spectroscopic data shows that these variations are driven by increasing electron density, which we attribute to a gradient in sulfur vacancy concentration. We propose that this gradient originates from directional precursor depletion near the nucleation site during growth. Together, these results establish a straightforward approach for producing $MoS_2$ layers with tunable, spatially controlled defect density, offering potential opportunities for optoelectronic and catalytic applications such as hydrogen evolution.

Experimental details:

Single-layer $MoS_2$ was synthesized in a CVD chamber using a standard down-flow configuration, with sulfur powder placed in the upper heating zone and $MoO_3$ powder in the lower zone, alongside the substrate positioned horizontally in a boat. Precursor amounts were set to m(S) = 240 mg, m($MoO_3$) = 4 mg, yielding large-area single-layer $MoS_2$ flakes. Optical microscopy was performed using a 100× objective. Additional morphological characterization was carried out by scanning electron microscopy (Verios 360L, Thermo Fisher Scientific) in ultra-high resolution (UHR) mode using secondary electron contrast, and by atomic force microscopy (AFM, Bruker Dimension Icon) in ScanAsyst mode, with subsequent data analysis performed in Gwyddion software. Spectroscopic mapping was performed using a 532 nm continuous-wave laser with 10 mW optical power, focused to a 0.5 μm spot size through a 100× objective. Photoluminescence and Raman spectra were acquired using 150 and 600 lines/mm gratings, respectively. Maps were recorded at 64×64 pixel resolution, with integration times of 0.1 s for photoluminescence and 2 s for Raman spectroscopy. Concerning power-dependent measurements, the laser power was adjusted using a rotatable gradual neutral density filter.

Acknowledgement

The authors gratefully acknowledge Prof. Yan-Wen Lan of the Department of Physics, National Taiwan Normal University, for providing laboratory facilities that made this study possible. Financial support from the National Science and Technology Council of Taiwan under Grant Nos. NSTC 114-2112-M-003-014 and NSTC 115-2112-M-003-007 is gratefully acknowledged. This work was also supported by the project "NEXT GENERATION BUT: Enhancing the Quality and Relevance of Education at BUT", funded as project No. CZ.02.02.XX/00/23_022/0009052 by Programme Johannes Amos Comenius. Additionally, CzechNanoLab Research Infrastructure (ID 90251), funded by MEYS CR, is gratefully acknowledged for the financial support of the measurements/sample fabrication. This work was also supported by the project CZ.02.01.01/00/22_008/0004594 TERAFIT. We acknowledge the support by the Czech Science Foundation (Grant No. 23-07617S).

Supplementary Information

In Supplementary Information, the gas flow direction is illustrated with optical microscope image across a large area of the sample surface.

Conflict of Interest

The authors have no conflict to disclose.

Author Contributions

M.K. performed the fabrication and sample analysis and wrote the manuscript, H.-S.C. and C.-Y.H. supported the experiments, Y.-R.C. contributed to the experiments, J.M. and M.B. contributed to the methodology and revised the manuscript, J.Š. assisted with experiments, T.Š. supervised the research project and W.-C.L. directed the entire study and revised the manuscript.

Data Availability

The datasets generated and analyzed during the current study are available in the Zenodo repository at https://doi.org/10.5281/zenodo.23949233.

Supplementary Information

Gas flow direction - OM image showing how the flakes change in the flow direction (white arrow). Up-stream flakes are bigger than down-stream. Gradient is oriented the same way as the gas flow. Defects are on the down-stream side.

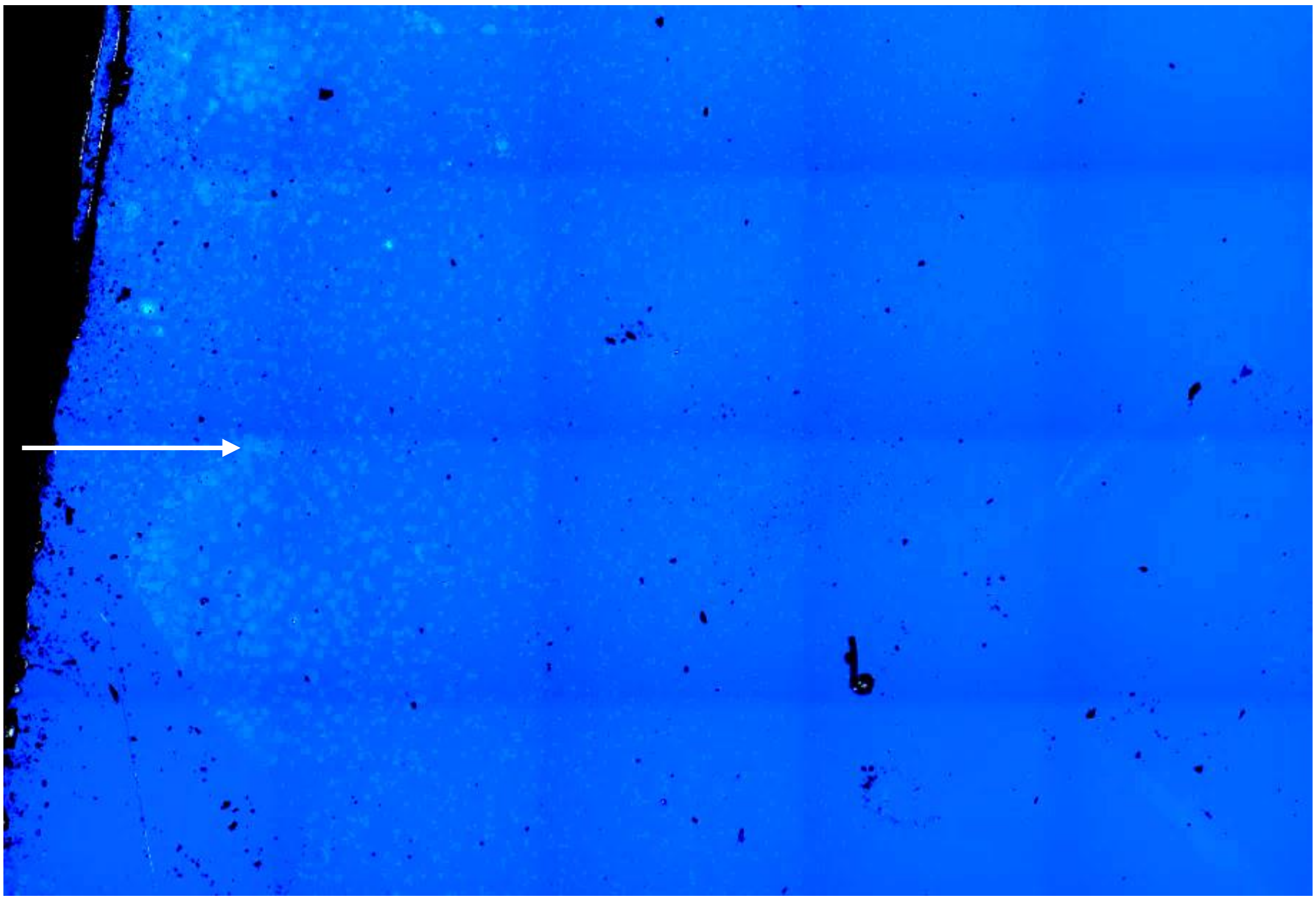

Fig. S1: Large area optical image of the sample with $MoS_2$ flakes